\documentclass[runningheads]{llncs}
\usepackage[T1]{fontenc}
\usepackage{graphicx}
\usepackage[T1]{fontenc}  
\usepackage{balance}
\usepackage{booktabs, multirow} 
\usepackage{soul}
\usepackage{changepage,threeparttable} 
\usepackage[colorinlistoftodos,prependcaption]{todonotes}
\usepackage{xspace}
\usepackage{listings}
\usepackage{comment}
\usepackage[scale=0.9]{sourcecodepro} 
\usepackage[subrefformat=parens]{subcaption}
\usepackage{algorithm}
\usepackage[noend]{algpseudocode}
\usepackage{graphicx}
\usepackage{siunitx}
\usepackage{textcomp}
\usepackage{xcolor}
\usepackage{makecell} 
\usepackage{amsfonts}
\usepackage{amsmath}
\usepackage{wrapfig}

\begin{document}
\title{Layer-wise CPU-GPU Scheduling for LLM Inference on Memory-Limited Consumer GPUs}

\titlerunning{A Layer-wise CPU-GPU LLM Scheduling for Consumer GPUs}
%
\author{Naoto Sugiura\inst{1} \and
Yuta Tokusashi\inst{2} \and
Hiroki Matsutani\inst{1}}
\authorrunning{N. Sugirua et al.}
%
\institute{Keio University, 3-14-1, Hiyoshi, Yokohama, Japan 223-8522\\ \email{\{nsugiura, matutani\}@arc.ics.keio.ac.jp} \and
SoftBank Corp., 1-7-1 Kaigan, Minato-ku, Tokyo, Japan, 105-7529\\
\email{yuta.tokusashi@g.softbank.co.jp}}
\maketitle              
%
\begin{abstract}
As large language model inference environments expand toward consumer devices, it remains challenging to efficiently run models that exceed GPU memory capacity in single-GPU settings. The latency of transferring weights that do not fit in GPU memory accounts for about 98\% of inference time and remains a major obstacle.
In this paper, we focus on the fact that during the decode phase, layer computation time on the CPU can be shorter than weight transfer time, presenting an opportunity to hide weight transfers by utilizing this computation time. 
Therefore, we propose a method to hide transfer latency by executing certain layers on the CPU during the decode phase while asynchronously transferring the next layer's weights to the GPU in the background of that computation. 
To realize this scheduling, we formulate the placement of CPU/GPU computation and weight transfers as a latency minimization problem and perform optimization using an approximation model with relaxed temporal constraints.
Evaluation results confirm that in a consumer desktop environment, our approach accelerates OPT-30B inference by 3.5$\times$ compared to HuggingFace Accelerate and 6.3$\times$ compared to DeepSpeed Zero-Inference.

\vspace{-2mm}
\end{abstract}


\section{Introduction}
\label{sec:intro}

As the social implementation of large language models (LLMs) accelerates, the environment for their inference is expanding from traditional, resource-rich cloud infrastructures to resource-constrained consumer devices, such as smartphones and personal computers (PCs). 
Market forecasts indicate that AI-powered PCs will account for 54.7\% of the PC market by 2026, while AI-powered desktops are expected to reach a 42.1\% share in the same year~\cite{gartner}. 
The growth of local execution on PCs enables reduced response times and power consumption while decreasing cloud dependency and enhancing data protection~\cite{gartner,xu2024device}. 
Moreover, another forecast suggests that the adoption of on-device AI, which does not rely entirely on the cloud, will be driven primarily by privacy concerns~\cite{bloomberg}. Beyond consumer PCs, LLMs are also anticipated to be integrated into fields such as robotics~\cite{zeng2023large} and autonomous driving~\cite{yang2023llm4drive}. Against this backdrop, LLM inference on edge and consumer devices is becoming increasingly important.

LLMs are utilized in a wide range of applications, including document generation, summarization, translation, and coding assistance. In particular, local inference on consumer devices is frequently employed for interactive applications, such as chatbots, which generate responses by sequentially producing tokens for each request. While some operational strategies aim for high throughput via batching~\cite{sheng2023flexgen,park2024improving}, this paper primarily assumes a local interactive setting with low concurrency. Key characteristics of this setting are that the effective batch size tends to be 1 and that minimizing latency, which directly affects user experience, is crucial~\cite{zhao2024hetegen}.

However, LLM inference demands substantial computational resources and memory capacity, presenting significant barriers to execution on resource-constrained consumer devices. For instance, the OPT-30B~\cite{zhang2022opt} model requires over 60~GiB of memory in FP16 format, which far exceeds the 32~GiB capacity of the RTX 5090, a state-of-the-art consumer-grade GPU. When LLM weights cannot fit within the GPU memory in a single-GPU environment, frameworks such as HuggingFace Accelerate~\cite{accel} typically place weights in host memory and transfer them to the GPU sequentially during computation. However, in this approach, weight transfer time dominates the total latency. For example, when using OPT-30B, transfer overhead accounts for approximately 98\% of the total execution time during the decoding phase, creating a severe performance bottleneck.

One direction for addressing this problem is model compression, such as quantization~\cite{xiao2023smoothquant}. However, some models still remain too large to fit in GPU memory, and aggressive compression can degrade accuracy. Another direction is offloaded inference, where weights and KV caches are placed on the host side. Prior work has proposed methods that reduce transfer stalls through scheduling that reuses weights across batches~\cite{sheng2023flexgen}, as well as through overlapping data transfers with computation and coordinating execution with the CPU~\cite{zhao2024hetegen,park2024improving,na2025flexinfer}. FlexGen is based on a simple overlap between GPU computation and data transfer, with only limited use of CPU computation~\cite{sheng2023flexgen}. As a result, prior work reports that exposed PCIe transfer time still accounts for approximately 96--98\% of the total execution time~\cite{na2025flexinfer}. Reducing latency in offloaded inference therefore hinges on how effectively transfer time can be hidden. Park and Egger~\cite{park2024improving} hide transfers by executing part of the intra-layer weights on the CPU and thereby accelerate throughput-oriented inference, whereas our paper differs in that it targets the reduction of exposed transfers during batch-size-1 decoding under offloading. HeteGen exploits CPU-side computation through CPU-GPU tensor parallelism, but its effectiveness depends on CPU computational capability. FlexInfer~\cite{na2025flexinfer} presents a framework that uses an estimator to select different execution policies for prefill and decode, but its choices are limited to CPU-only execution, GPU execution with transfers from host memory, and a static split called SplitGen, in which only the portion of weights that fits on the GPU is executed on the GPU and the remainder is executed on the CPU. It does not address the direction of hiding transfers behind CPU computation.

In this paper, we propose layer placement and scheduling techniques to minimize exposed transfer time for offload inference in a single-GPU environment, specifically targeting decoding with a batch size of 1. Our key observation is that the per-layer computation time on the CPU may be shorter than the weight transfer time during the decoding process.
In this case, CPU computation is not merely an alternative execution target but can serve as a planned computation window for hiding transfers. Based on this observation, we assign each layer to one of three modes: execution on the CPU, residency on the GPU, or placement in host memory with transfer to the GPU at execution time followed by GPU computation, which we call the stream mode. We then construct a latency model based on the computation time and transfer time of each mode. We further introduce an approximation model that relaxes the temporal constraints of transfers and reduce the complex scheduling optimization to a ratio determination problem, thereby synthesizing an execution plan from offline measurements, namely per-layer computation time and transfer time. At runtime, transfer for stream execution blocks is hidden behind CPU computation blocks, thereby minimizing inference latency.

In summary, this work offers the following contributions:
\begin{itemize}
\vspace{-1mm}
    \item We conduct an extensive analysis of LLM inference in consumer-grade desktop environments and quantitatively characterize the workload properties.
    \item We formulate the latency minimization problem for LLM inference in single-GPU environments as an optimization problem involving the placement of GPU residency, CPU computation, and stream computation.
    \item To accelerate LLM inference on consumer desktops, we propose a method that minimizes latency by efficiently overlapping CPU computation with weight transfers.
    \item Our evaluation demonstrates that for OPT-30B, the proposed method achieves an inference speedup of 3.5$\times$ compared to Hugging Face Accelerate~\cite{accel} and 6.3$\times$ compared to DeepSpeed ZeRO-Inference~\cite{aminabadi2022deepspeed}.
\end{itemize}

\section{Background and motivation}
\label{sec:back}
\subsection{LLM inference}

LLMs are composed of multiple Transformer layers~\cite{vaswani2017attention} and generate text by repeatedly predicting the next token. This inference process can be broadly divided into two stages, prefill and decode. First, the stage that processes the entire input prompt at once is called prefill. In this stage, the Key and Value tensors produced by the attention computation in each layer are accumulated as a KV cache, and the first token is predicted. After that, the stage that generates new tokens one by one while referring to and updating the KV cache is called decode. Whereas prefill benefits from the parallel processing capability of GPUs, decode is inherently sequential, and the latency of each step directly affects user-perceived response time.

Within each layer, multi-head attention (MHA) and a multi-layer perceptron (MLP) are mainly executed in sequence. MHA is responsible for extracting contextual information by combining Values according to the similarity between Queries and Keys. In decode, the KV cache is reused to avoid recomputing past Keys and Values, but the amount of memory access in attention increases as the sequence length grows. The subsequent MLP applies two linear transformations and an activation function to the output of attention, enabling more advanced information transformation.

\subsection{Performance analysis}
\label{ssec:pref}

\begin{figure}[t]
    \centering
    \begin{minipage}{0.49\columnwidth}
        \includegraphics[width=\linewidth]{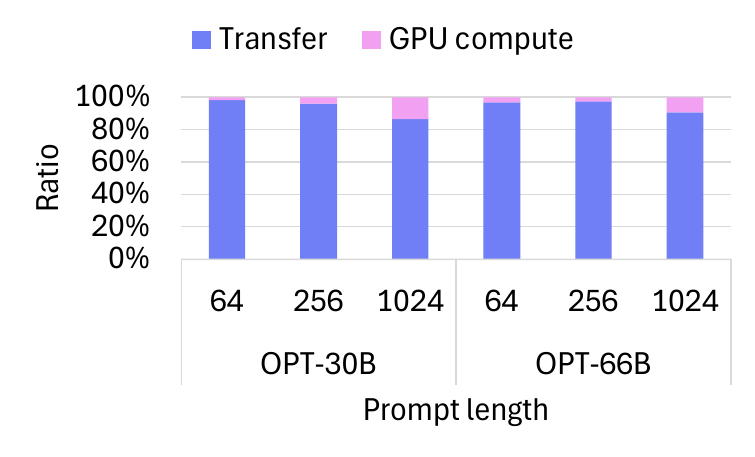}
        \vspace{-7mm}
        \subcaption{Prefill.}
        \label{fig:prefill}
    \end{minipage}
    \hfill 
    \begin{minipage}{0.49\columnwidth}
        \includegraphics[width=\linewidth]{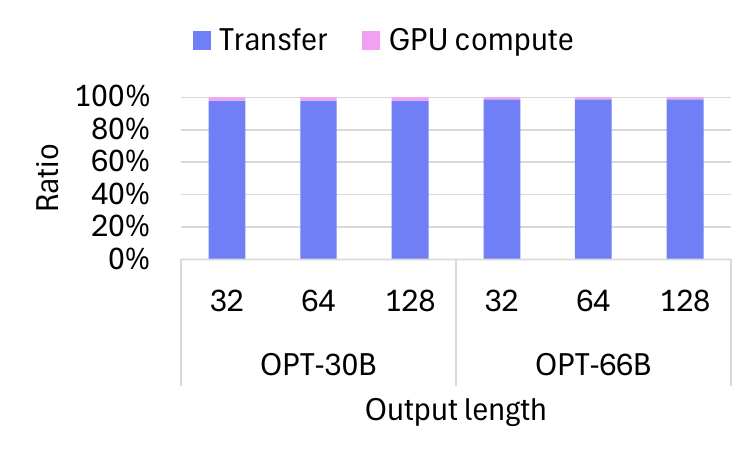}
        \vspace{-7mm}
        \subcaption{Decode (Prompt length: 256).}
        \label{fig:cpu_overhead}
    \end{minipage}
    \vspace{-3mm}
    \caption{Execution time breakdown per phase. (Batch size: 1, Type: FP16)}
    \label{fig:phase}
    \vspace{-4mm}
\end{figure}

\begin{figure}[t]
    \centering
    \begin{minipage}{0.49\columnwidth}
        \includegraphics[width=\linewidth]{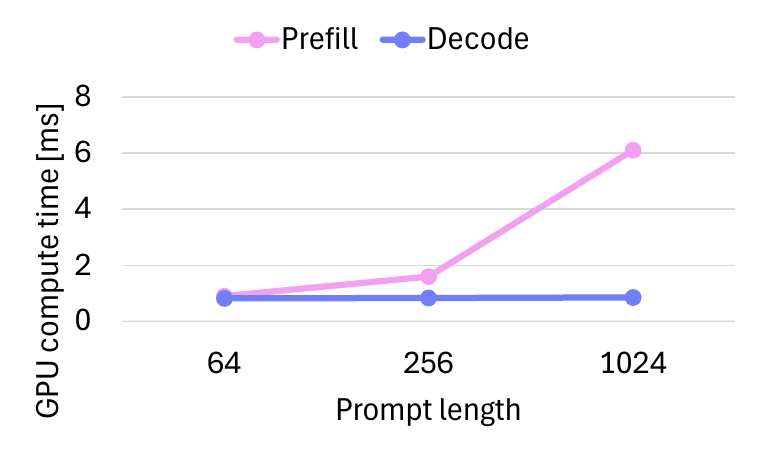}
        \vspace{-7mm}
        \subcaption{OPT-30B.}
        \label{fig:opt30b-gpu}
    \end{minipage}
    \hfill 
    \begin{minipage}{0.49\columnwidth}
        \includegraphics[width=\linewidth]{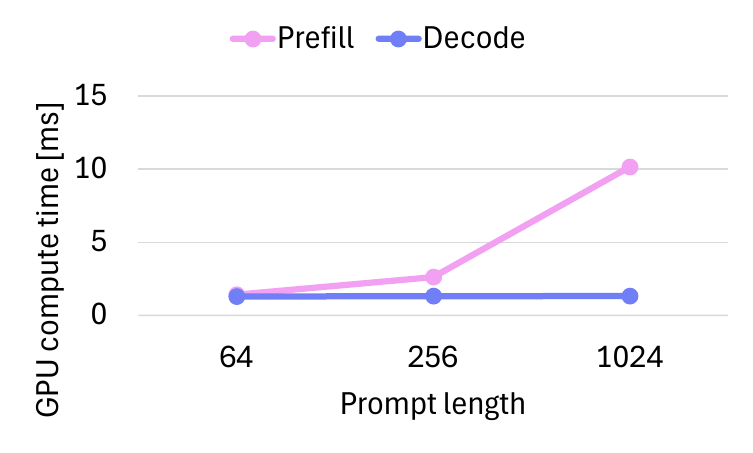}
        \vspace{-7mm}
        \subcaption{OPT-66B.}
        \label{fig:opt66b-gpu}
    \end{minipage}
    \vspace{-3mm}
    \caption{GPU computation time per layer for each phase.}
    \label{fig:gpu}
    \vspace{-4mm}
\end{figure}

\begin{figure}[t]
    \centering

    \begin{minipage}{0.49\linewidth}
        \centering
        \includegraphics[width=\linewidth]{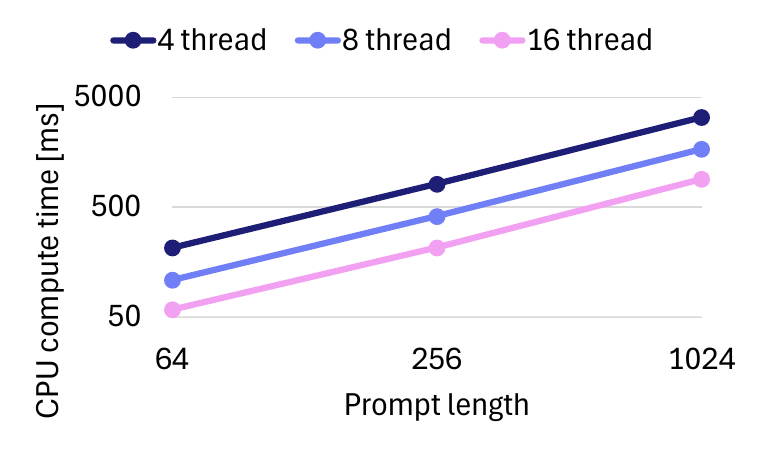}
        \vspace{-7mm}
        \subcaption{OPT-30B. Prefill.}
        \label{fig:opt30b-prefill}
    \end{minipage}
    \hfill
    \begin{minipage}{0.49\linewidth}
        \centering
        \includegraphics[width=\linewidth]{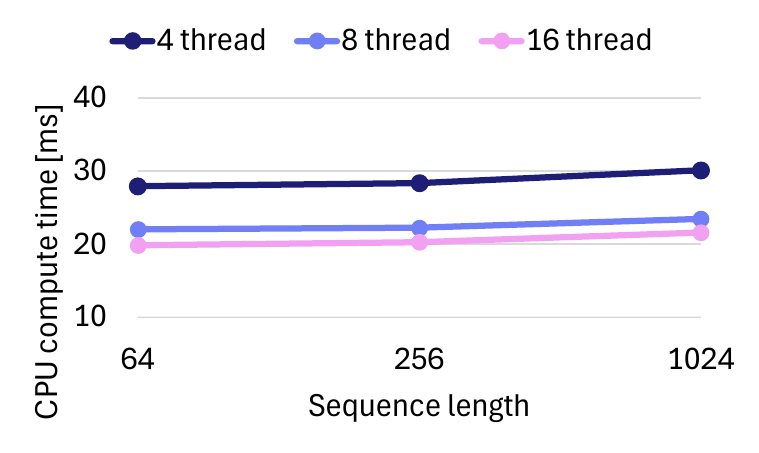}
        \vspace{-7mm}
        \subcaption{OPT-30B. Decode.}
        \label{fig:opt30b-decode}
    \end{minipage}

    \vspace{1mm}

    \begin{minipage}{0.49\linewidth}
        \centering
        \includegraphics[width=\linewidth]{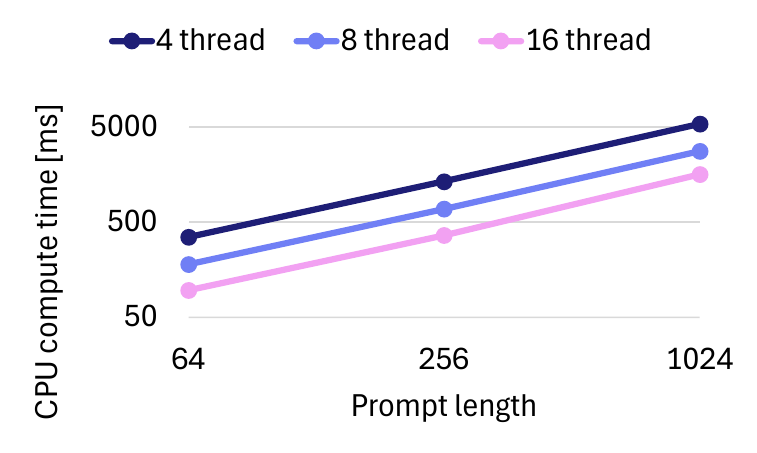}
        \vspace{-7mm}
        \subcaption{OPT-66B. Prefill.}
        \label{fig:opt66b-prefill}
    \end{minipage}
    \hfill
    \begin{minipage}{0.49\linewidth}
        \centering
        \includegraphics[width=\linewidth]{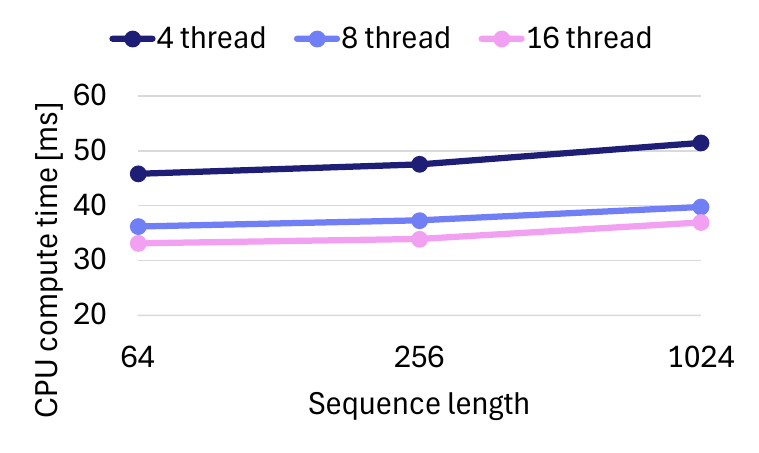}
        \vspace{-7mm}
        \subcaption{OPT-66B. Decode.}
        \label{fig:opt66b-decode}
    \end{minipage}
    \vspace{-3mm}
    \caption{CPU computation time per layer for each phase.}
    \label{fig:cpu}
    \vspace{-5mm}
\end{figure}

\begin{table}[t]
  \centering
  \caption{Weight transfer time per layer over PCIe for OPT models (Pinned vs. Pageable memory).}
  \label{tab:pci_transfer_time}
  \begin{tabular}{lrr}
    \hline
    \textbf{DMA Type} & \textbf{OPT-30B} & \textbf{OPT-66B} \\ \hline
    Pinned      & 28.2 ms  & 47 ms  \\
    Pageable    & 79.8 ms  & 129.9 ms \\ \hline
  \end{tabular}
  \vspace{-4mm}
\end{table}

In LLM inference, workload characteristics vary with runtime parameters such as the phase and the input and output lengths. In this section, we analyze the performance of LLM inference in a single-GPU setting where the model weights do not fully fit in GPU memory. The hardware configuration follows Section~\ref{ssec:eval_setup}. In this paper, we assume batch size 1 as the primary use case for LLM inference on consumer PCs.
HuggingFace Accelerate~\cite{accel} places the portion of weights that fits in GPU memory on the GPU and keeps the remaining weights in host memory, transferring them to the GPU on demand during execution. In this approach, computation and transfer are serialized, so weight transfer time directly contributes to end-to-end latency.
As shown in Figure~\ref{fig:phase}, transfers account for most of the latency, indicating that host-to-GPU transfer becomes the bottleneck when the full set of weights cannot reside in GPU memory on a single GPU. During prefill, transfers account for 86--98\% of the execution time for OPT-30B. As the prompt length increases, the transfer fraction decreases, but it remains substantial. During decode, transfers account for 98\% of the execution time and remain constant regardless of the generation length. OPT-66B exhibits the same trend.

Because host-to-GPU transfers are slower than GPU computation, recent work has proposed cooperative execution with CPU computation~\cite{park2024improving,zhao2024hetegen,na2025flexinfer}. However, some studies primarily evaluate server-class CPUs equipped with matrix extensions such as Intel Advanced Matrix Extensions (AMX). While some evaluations include environments with desktop CPUs, there are limited reports that quantify the relationship among GPU computation, CPU computation, and transfer time in the batch-size-1 decode setting that we primarily target. Therefore, we conduct layer-level profiling in a consumer desktop environment.

Figure~\ref{fig:gpu} and Figure~\ref{fig:cpu} show per-layer computation times on the GPU and CPU, respectively. For the CPU, we also report results across different thread counts. Table~\ref{tab:pci_transfer_time} reports transfer time from host memory to GPU memory.
Comparing GPU and CPU computation times, the GPU is more than 100$\times$ faster. This gap is especially pronounced in the compute-intensive prefill phase, where the GPU can be more than 1000$\times$ faster. Since prefill is compute-intensive, CPU execution can be accelerated by increasing the number of threads. However, in many cases, CPU execution remains slower than the sum of transfer time and GPU computation time, indicating that transferring weights to the GPU and computing there is more efficient than computing on the CPU. In decode, increasing the thread count provides limited benefit. Nevertheless, comparing transfer time and CPU time shows that CPU computation is faster than transfer in the decode phase. This indicates room for speedup by overlapping CPU computation with transfers. OPT-66B has more parameters per layer than OPT-30B, which increases latency, but the overall trends remain the same.

\subsection{Motivation}

GPUs accelerate LLM inference through their high matrix computation capability. However, in a single-GPU environment where all model weights cannot reside in GPU memory, a practical design is to offload weights to host memory and transfer them to the GPU on demand. In this setting, inference latency is dominated by weight transfer time.
As shown by the measurements in Section~\ref{ssec:pref}, during decode, per-layer CPU computation time can be shorter than weight transfer time, which suggests an opportunity to use CPU computation as a planned computation window for hiding transfers.

However, to actually reduce exposed transfers, it is necessary to jointly decide at the layer level which layers should be assigned to the CPU to create hiding windows and when the weights of other layers should be transferred. As discussed in Section~\ref{sec:intro}, this design space has not been sufficiently explored in prior work.
In this paper, we aim to minimize the inference latency of one decode step by selecting, for each layer, among CPU execution, GPU residency, and transferred GPU execution, and by using the layers executed on the CPU as hiding intervals for weight transfers. However, how to optimally assign these placements to individual layers is not straightforward.

\section{Problem formulation}
\label{sec:formula}


\subsection{Analytical model}

In a single LLM decode step (one-token generation), each layer $i \in \{1, \dots, N\}$ is executed serially. We formulate the latency by focusing on the fact that each layer can adopt one of the following three execution modes.
\begin{itemize}
\vspace{-1mm}
   \item \textbf{cpu}: Execute on the CPU. The computation time of layer $i$ is $c_i$.
   \item \textbf{resident}: Execute on the GPU, and the weights are resident on the GPU. The computation time of layer $i$ is $g_i$, and no weight transfer is required.
   \item \textbf{stream}: Execute on the GPU, but the weights are transferred on demand. The computation time of layer $i$ is $g_i$, and a transfer time $w_i$ is required.
\end{itemize}
The latency is determined once the execution mode $m_i \in \{\mathrm{cpu}, \mathrm{resident}, \mathrm{stream}\}$ is selected for each layer. 
Since both resident and stream modes utilize the GPU for computation, we define an indicator variable $u_i$ that represents whether layer $i$ is computed on the GPU as $u_i = \mathbf{1}[m_i \neq \mathrm{cpu}]$.
Because the layers are processed serially, the total latency $L$ for one decode step is expressed as the sum of the following four terms.

\begin{enumerate}
\vspace{-1mm}
   \item Total computation time of layers executed on the CPU: $T_{\mathrm{cpu}}(m) = \sum_{i=1}^{N} (1 - u_i) c_i$
   \item Total computation time of layers executed on the GPU: $T_{\mathrm{gpu}}(m) = \sum_{i=1}^{N} u_i g_i$
   \item Exposed transfer time (stall): Weight transfers for stream layers can proceed in parallel with computation, but any portion that cannot be fully hidden by computation appears as waiting time. We define this as $T_{\mathrm{stall}}(m)\ge 0$.
   \item Activation transfer (CPU-GPU switching): Each time the execution device switches between adjacent layers, an activation transfer cost $a$ is incurred. Let the number of switches be $B(m)=\sum_{i=1}^{N-1} |u_i-u_{i+1}|$, then the activation transfer time is $aB(m)$.
\end{enumerate}

Based on these components, the latency $L$ for a single decode step is represented by the following equation.
\begin{equation}
L(m) = T_{\mathrm{cpu}}(m) + T_{\mathrm{gpu}}(m) + T_{\mathrm{stall}}(m) + aB(m)
\end{equation}
The objective of this paper is to select $m$ to minimize $L(m)$ while satisfying the GPU memory capacity constraints.

\subsection{Approximation via transfer-computation balancing}

The stall term $T_{\mathrm{stall}}(m)$ 
depends on detailed temporal conditions, including when each transfer starts, which computation interval it overlaps with, and whether the transfer finishes before the corresponding layer is executed. Directly modeling these conditions leads to a complicated analysis. Therefore, rather than tracking the fine-grained timing of each individual transfer, we introduce an approximation model that focuses on the relation between the total time required for transfers and the total amount of computation time available to hide them. In this approximation, transfers are treated as if they can be flexibly assigned across computation intervals, and the individual deadline constraints of each transfer are relaxed.
Under this approximation, the stall term is interpreted as the portion of the total transfer time that remains exposed even after overlap with computation. We also assume that CPU computation and weight transfer to the GPU can proceed independently and can fully overlap.

First, the total transfer time required for layers executed in stream mode is expressed as $W(m)=\sum_{i=1}^{N}[m_i=\mathrm{stream}]\, w_i$.
Similarly, the total amount of computation time that can potentially hide transfers is expressed as $C(m)=T_{\mathrm{cpu}}(m)+T_{\mathrm{gpu}}(m)$.
In the approximation model, we assume that the exposed stall is the portion of the total transfer time $W(m)$ that cannot be hidden within the computation time $C(m)$, which yields
\begin{equation}
T_{\mathrm{stall}}(m)\approx \max(0,\;W(m)-C(m)).
\end{equation}
Furthermore, since $a$ is sufficiently small compared with the other terms, we ignore it in the following analysis. In our experimental environment, $a$ is only 0.005~ms. The latency in the previous subsection can then be approximated as
\begin{equation}
\begin{aligned}
L(m) &\approx C(m) + \max(0,\;W(m)-C(m)) \\
     &= \max(C(m), W(m)).
\end{aligned}
\end{equation}
That is, the latency is determined by whichever is larger, the total computation time $C(m)$ or the total transfer time $W(m)$.
In general, in offloaded inference on a single GPU, transfer time is much larger than computation time, that is, $w_i \gg g_i$. Therefore, in the standard approach that executes all layers on the GPU, we typically have $W(m) > C(m)$, and transfer becomes the bottleneck. To mitigate this bottleneck, it is effective to intentionally increase $C(m)$ by executing some layers on the slower CPU, thereby hiding the transfer cost of the remaining stream layers within that computation time.

Based on this idea, we next consider the optimal execution block while respecting the GPU memory capacity limit. To efficiently overlap CPU computation with weight transfer, it is desirable to form an execution block in which the weights of $y$ stream layers are transferred asynchronously to the GPU while $x$ layers are computed on the CPU. If the computation time is approximately uniform across layers, the increase in total computation time caused by assigning a layer to the CPU can be approximated by the average $\bar{\Delta}=\mathbb{E}[c_i-g_i]$ and the increase in total transfer time caused by assigning a layer to stream mode can be approximated by the average $\bar{w}=\mathbb{E}[w_i]$.
Here, the GPU computation time $g_i$ is much smaller than both the CPU computation time $c_i$ and the transfer time $w_i$, that is, $g_i \ll c_i, w_i$. Thus, in practice, we can approximate $C(m) \approx T_{\mathrm{cpu}}(m)$. Moreover, we have $\bar{\Delta} \approx \mathbb{E}[c_i] = \bar{c}$.
Under these approximations, minimizing $\max(C(m), W(m))$ becomes equivalent to balancing $T_{\mathrm{cpu}}(m)$ and $W(m)$. Hence, a desirable ratio satisfies $\bar{c}x \approx \bar{w}y$.
Therefore, under the integer constraint, we solve
\begin{equation}
\label{eq:min}
\min_{x,y\in\mathbb{Z}}
\left|\bar{c}\,x-\bar{w}\,y\right|
\end{equation}
and use the resulting ratio $x:y$ to allocate non-resident layers between cpu and stream modes.

\section{Proposed method}
\label{ssec:overview}

\begin{figure}[t]
  \centering
  \includegraphics[width=\columnwidth]{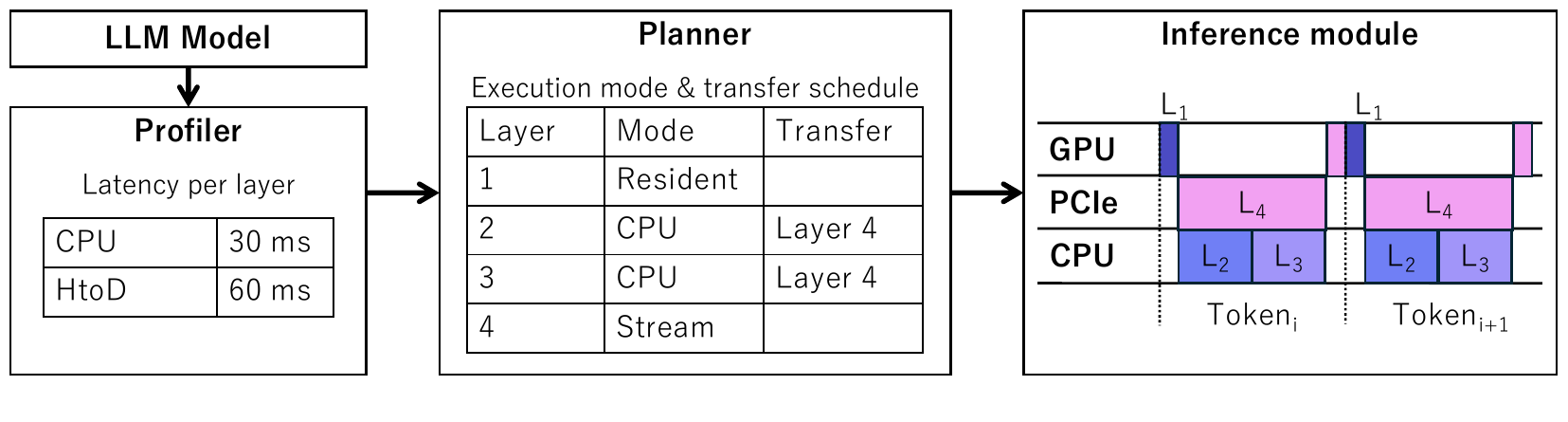}
  \vspace{-9mm}
  \caption{Overview of proposed method.}
  \label{fig:method}
  \vspace{-4mm}
\end{figure}

In this section, we describe how to construct an execution plan based on the formulation in Section~\ref{sec:formula} and how to run inference accordingly. Figure~\ref{fig:method} provides an overview. First, our method obtains per-layer CPU computation time and transfer time via offline profiling. Then, it determines the execution mode for each layer, and the inference module runs inference following the resulting plan.

\subsection{Execution plan construction}
\label{ssec:structure}

\begin{figure}[t]
  \centering
  \includegraphics[width=\columnwidth]{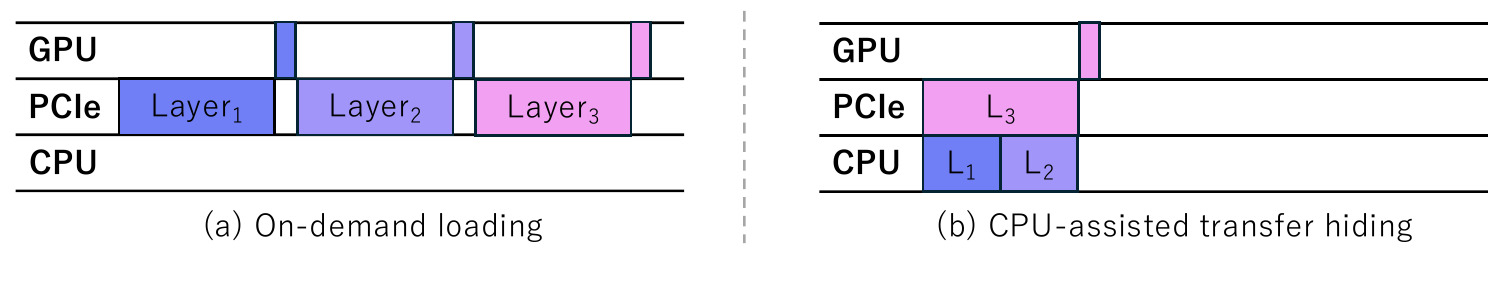}
  \vspace{-12mm}
  \caption{Comparison of on-demand loading and CPU-assisted transfer hiding}
  \label{fig:workflow}
  \vspace{-6mm}
\end{figure}

We next construct the execution plan by assigning each layer to resident, CPU, or stream execution.
To obtain $x$ and $y$ that satisfy Eq.~\ref{eq:min}, the average computation time per layer $\bar{c}$ and the average transfer time per layer $\bar{w}$ are required. 
In our method, we use values collected through offline profiling to compute the numbers of layers $x$ and $y$ that form the optimal execution block.
Using these profiled values, Algorithm~\ref{alg:min_cx_wy} performs a continued-fraction search to find integer values of $x$ and $y$ that minimize $|\bar{c}x-\bar{w}y|$ under the constraint $x+y\le N$.

Next, we determine the number of resident layers, denoted by $n_{resident}$. Let $K$ be the effective GPU memory capacity, that is, the maximum number of layers that can be used for weight placement after excluding the space required for the KV cache and temporary activation buffers. In the inference process of our method, an additional temporary space is required to transfer and retain the weights of $y$ layers as a batch, which we call the stream buffer. Therefore, to maximize the number of resident layers while satisfying the GPU memory capacity constraint, $n_{resident}$ is uniquely determined as $n_{resident}=K-y$, which is obtained by subtracting the $y$ layers reserved for the stream buffer from the effective capacity $K$.

Among the total $N$ layers, the remaining $N-n_{resident}$ layers that cannot reside on the GPU are partitioned into blocks with the computed ratio of $x$ to $y$. At runtime, these blocks are scheduled so that CPU computation and stream execution are processed alternately and in parallel. As shown in Figure~\ref{fig:workflow}, this reduces inference latency compared with an on-demand loading scheme. As a result, the method fully utilizes precious GPU memory for resident layers while using only the minimum required stream buffer, enabling inference that completely hides transfer latency.

\begin{algorithm}[t]
\caption{Continued-fraction search under a sum constraint}
\label{alg:min_cx_wy}
\begin{algorithmic}[1]
\Require Positive integers $\bar{c},\bar{w},n$
\Ensure Integer pair $(x,y)\neq(0,0)$ minimizing $|\bar{c}x-\bar{w}y|$

\State $(x^\star,y^\star) \gets (\bar{c} < \bar{w}) ? (1, 0) : (0, 1)$
\State $\delta^\star \gets |\bar{c} x^\star - \bar{w} y^\star|$

\State $(p_{prev}, q_{prev}) \gets (1,0)$, \ $(p_{curr}, q_{curr}) \gets (\lfloor \bar{c}/\bar{w} \rfloor, 1), \ (u,v) \gets (\bar{c},\bar{w})$

\While{\textbf{true}}
    \If{$q_{curr} + p_{curr} \le n$}
        \State $\delta \gets |\bar{c} q_{curr} - \bar{w} p_{curr}|$
        \If{$\delta < \delta^\star$}
            \State $(x^\star,y^\star,\delta^\star) \gets (q_{curr}, p_{curr}, \delta)$
        \EndIf
        
        \If{$\delta^\star = 0$}
            \State \textbf{break}
        \EndIf
        
    \Else
        \State $k \gets \left\lfloor \dfrac{n-(q_{prev}+p_{prev})}{q_{curr}+p_{curr}} \right\rfloor$
        \If{$k > 0$}
            \State $(\hat{x},\hat{y}) \gets (q_{prev}+k q_{curr},\; p_{prev}+k p_{curr})$
            \State $\hat{\delta} \gets |\bar{c}\hat{x}-\bar{w}\hat{y}|$
            \If{$\hat{\delta} < \delta^\star$}
                \State $(x^\star,y^\star,\delta^\star) \gets (\hat{x}, \hat{y}, \hat{\delta})$
            \EndIf
        \EndIf
        \State \textbf{break}
    \EndIf

    \State $r \gets u \bmod v$
    \If{$r = 0$}
        \State \textbf{break}
    \EndIf
    \State $(u,v) \gets (v,r)$
    \State $\alpha \gets \left\lfloor u/v \right\rfloor$
    \State $(p_{next}, q_{next}) \gets (\alpha p_{curr} + p_{prev},\; \alpha q_{curr} + q_{prev})$
    \State $(p_{prev}, q_{prev}) \gets (p_{curr}, q_{curr})$
    \State $(p_{curr}, q_{curr}) \gets (p_{next}, q_{next})$
\EndWhile

\State \Return $(x^\star,y^\star)$
\end{algorithmic}
\end{algorithm}

\subsection{Inference procedure}

We then describe how inference is executed under this plan.
At runtime, the resident layers are first transferred to the GPU. During inference, layers are processed in order, and each layer is executed on either the CPU or the GPU according to the plan. For stream layers that are executed on the GPU, asynchronous transfer is initiated immediately before the computation of the first CPU layer in the execution block. At this point, the transfer proceeds in parallel with the computation of the CPU layers. After the transfer completes, the location of the weights referenced by the corresponding layer is switched to the GPU side, and the computation is performed when that layer is reached. Activations are moved to the required device only when the computation device changes between adjacent layers.


In prefill, the KV cache for past tokens does not yet exist, so auxiliary inputs such as positional information and attention masks are prepared only once at the beginning of the forward pass and then reused across layers. This reuse is possible because these auxiliary inputs do not change across layers for the same input sequence, and it avoids the cost of regenerating them for each layer. As shown by the analysis in Section~\ref{ssec:pref}, CPU computation is slower in the prefill phase, so transferring the weights to the GPU and executing there is faster. Therefore, in the prefill phase, even layers assigned to the CPU are transferred to the GPU and executed there.

In decode, tokens are generated one by one while retaining the KV cache for past tokens. Since the plan divides layers between the CPU and the GPU, the KV cache is also placed according to the execution device of each layer. Specifically, The KV cache of CPU-executed layers is placed on the host side, while that of all other layers is placed on the GPU. With FP16 and batch size 1, the maximum KV cache size per layer is 56~MiB for OPT-30B and 72~MiB for OPT-66B, so its memory impact is small compared with that of the weights.
After that, the layers are processed sequentially at each step, and inference proceeds in the same manner while inserting transfers of weights and activations when needed. At each step, the completion of weight transfers and computation is synchronized with CUDA events, which preserves the dependencies among transfer, computation, switching of weight references, and reuse of temporary buffers.

\subsection{Implementation details}

Our method is implemented on top of HuggingFace Transformers~\cite{wolf2019huggingface}. To minimize transfer latency, we use pinned memory. The weights of stream-targeted layers are packed into contiguous memory regions 
and kept on the host side in page-locked memory. 
These weights are then transferred asynchronously to a temporary buffer on the GPU, after which the storage location of the weights referenced by the target layer is switched to the GPU side and computation is performed. After computation, the reference is switched back to the host side, and the temporary buffer on the GPU is overwritten and reused. Therefore, along this path, no weight transfer is performed from the GPU back to the host.

\section{Evaluation}

\subsection{Setup}
\label{ssec:eval_setup}


Because our method does not use compression or approximation and therefore causes no accuracy degradation, we evaluate only performance. As performance metrics, we use end-to-end latency and energy consumption.
The evaluation is conducted on a system equipped with an NVIDIA RTX 5090 with 32~GiB of memory and an Intel Core Ultra 7 265F @ 2.40 GHz with 128~GiB of memory. The GPU and the host are connected via PCIe Gen5 $\times$16. 
Experiments are conducted using NVIDIA CUDA 12.4.131.
CPU computation is configured to use 8 threads. Although the Core Ultra 7 includes an NPU, it is not used in this paper.
We use OPT-30B and OPT-66B~\cite{zhang2022opt} and Llama2-70B~\cite{touvron2023llama} as the target models. FP16 is used as the data type. We evaluate the models with prompt lengths of 256 and 1024, while fixing the decode length to 32.
As baselines, we use Hugging Face Accelerate~\cite{accel}, DeepSpeed Zero-Inference~\cite{aminabadi2022deepspeed}, and two policies from FlexInfer, namely CPU-only and SplitGen. CPU-only executes all computation on the CPU. SplitGen executes the portion of weights that fit in GPU memory on the GPU, while executing the remaining portion on the CPU.

\subsection{Inference performance}

\begin{figure*}[t]
    \centering
    \begin{minipage}{0.32\textwidth}
        \includegraphics[width=\linewidth]{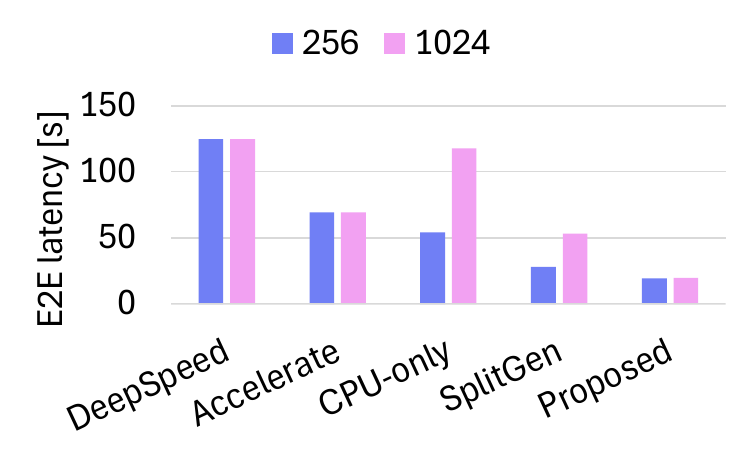}
        \vspace{-5mm}
        \subcaption{OPT-30B.}
        \label{fig:ex1-boxplot}
    \end{minipage}
    \hfill 
    \begin{minipage}{0.32\textwidth}
        \includegraphics[width=\linewidth]{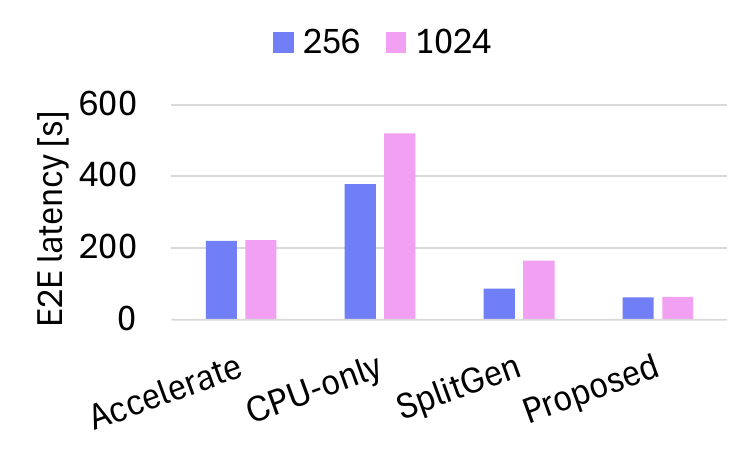}
        \vspace{-5mm}
        \subcaption{OPT-66B.}
        \label{fig:ex2-update}
    \end{minipage}
    \hfill 
    \begin{minipage}{0.32\textwidth}
        \includegraphics[width=\linewidth]{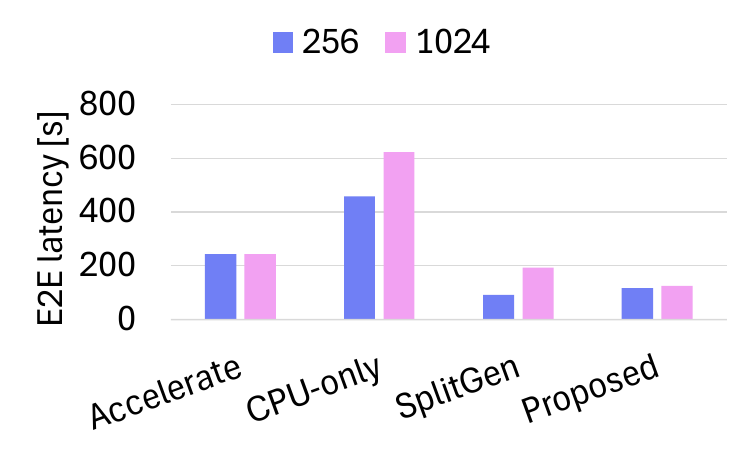}
        \vspace{-5mm}
        \subcaption{Llama2-70B.}
        \label{fig:ex2-scan}
    \end{minipage}
    \vspace{-3mm}
    \caption{End-to-end latency comparison with prompt lengths of 256 and 1024. DeepSpeed is omitted for OPT-66B and Llama2-70B because of out of memory.}
    \label{fig:ex1-e2e}
    \vspace{-4mm}
\end{figure*}

Figure~\ref{fig:ex1-e2e} shows the end-to-end latency of each method for OPT-30B, OPT-66B, and Llama2-70B with prompt lengths of 256 and 1024. 
For OPT-30B with a 256-token prompt, the proposed method is 3.5$\times$, 6.3$\times$, 2.8$\times$, and 1.5$\times$ faster than Accelerate, DeepSpeed, CPU-only, and SplitGen, respectively.
For OPT-66B, the proposed method is 3.5$\times$ faster than Accelerate, 6.1$\times$ faster than CPU-only, and 1.4$\times$ faster than SplitGen. 
For Llama2-70B, the proposed method incurs 22\% higher latency than SplitGen when the prompt length is 256.
This is because Llama2-70B itself is large, and pinned memory further increases pressure on host memory.
On-demand loading methods such as Accelerate and DeepSpeed process all layers on the GPU, so latency varies little with changes in prompt length.
Similarly, because the proposed method also uses on-demand loading for the prefill stage, the difference in latency due to prompt length is small.
In contrast, CPU-only and SplitGen perform the prefill on the CPU, which leads to a significant difference based on prompt length. In the CPU-only method, latency increases as the model size grows because the computational load increases.
While transfer time is a major bottleneck for Accelerate and DeepSpeed, the proposed method achieves higher speeds by utilizing CPU computational resources and performing computation in parallel with data transfer. 

\begin{wrapfigure}[10]{r}{0.28\textwidth}
  \vspace{-2.6\baselineskip}
  \centering
  \includegraphics[width=0.26\textwidth]{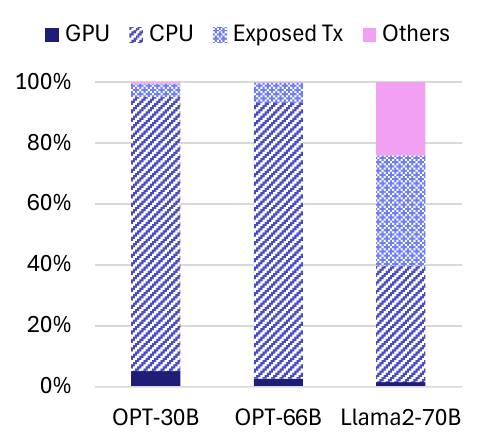}
  \vspace{-2mm}
  \caption{Breakdown of execution time. Tx denotes transfer.}
  \label{fig:breakdown}
\end{wrapfigure}

Figure~\ref{fig:breakdown} shows the execution time breakdown of the proposed method with a 256-token prompt. For OPT-30B and OPT-66B, the ratio of exposed transfers is minimal, as 99.6\% and 99.8\% of the transfers are hidden by computation, respectively. In contrast, almost all transfers are exposed for Llama2-70B. The plan is made using average computation and transfer times. However, host memory pressure makes these times unstable, so the execution does not follow the plan as expected. A potential countermeasure is to adopt worst-case measurements during the profiling phase. 
The major overhead of our method is execution plan construction. However, this takes only 3.2 seconds, and because the plan is reusable, its impact on overall performance is negligible.

\subsection{Energy consumption}

\begin{wrapfigure}[5]{r}{0.37\textwidth}
  \vspace{-2.6\baselineskip}
  \centering
  \includegraphics[width=0.35\textwidth]{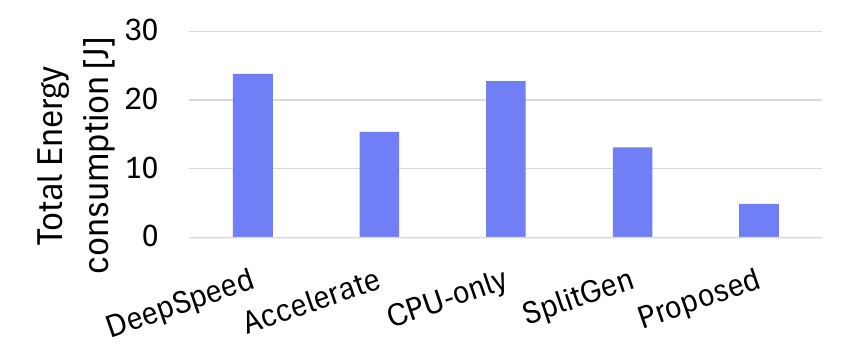}
  \vspace{-2mm}
  \caption{Energy consumption.}
  \label{fig:energy}
\end{wrapfigure}

The total energy consumption was calculated by integrating the power values sampled at 1~Hz using a watt-hour meter during the inference period. Figure~\ref{fig:energy} shows the energy consumption for a single inference of OPT-30B with a 1024-token prompt and 32-token output.
The proposed method achieves the lowest total energy consumption, outperforming DeepSpeed, Accelerate, CPU-only, and SplitGen by 4.9$\times$, 3.1$\times$, 4.6$\times$, and 2.7$\times$, respectively, primarily due to its shorter inference time.
The proposed method and SplitGen, both of which perform computation on both the CPU and GPU, exhibit higher average power consumption, at 273~W and 262 W, respectively. In contrast, the average power consumption is 219~W for Accelerate, 199~W for DeepSpeed, and 193~W for CPU-only. However, because DeepSpeed and CPU-only have longer execution times, their total energy consumption is higher.
\section{Related work}



FlexGen~\cite{sheng2023flexgen} proposes a zig-zag scheduling that reuses weights across batches to maximize throughput.
Park and Egger~\cite{park2024improving} reduce single-layer latency by using dynamic programming to split each layer into three segments, with the first and third executed on the CPU and the second on the GPU.
These approaches target batch-oriented workloads and focus on throughput-oriented inference acceleration. 
HeteGen~\cite{zhao2024hetegen} overlaps CPU computation, transfer, and GPU computation through heterogeneous tensor parallelism, but relies on substantial CPU compute capability.
FlexInfer~\cite{na2025flexinfer} adapts execution to runtime conditions by choosing among CPU-only, on-demand GPU execution, and SplitGen, but does not use CPU execution as a transfer-hiding window. 
Relatedly, KVPR~\cite{jiang2025kvpr} and ShadowKV~\cite{sun2024shadowkv} target KV-cache offloading for long-context inference.
In contrast, our work targets batch size 1 decode with weight offloading and minimizes exposed transfer time through layer-wise placement into CPU, resident GPU, and stream modes.

\section{Conclusion}

In this paper, we aimed to minimize LLM inference latency in resource-constrained single-GPU environments and proposed an optimization method for placing GPU computation, CPU computation, and stream-based computation.
In conventional offloading methods, weight transfer over PCIe has been a bottleneck. Our method addresses this issue by executing some layers on the CPU and asynchronously transferring weights during that computation time, thereby realizing execution blocks that hide transfer latency.
Our evaluation on a consumer desktop equipped with an RTX 5090 showed that, for OPT-30B, our method achieved a 3.5$\times$ speedup and 3.1$\times$ lower energy consumption than HuggingFace Accelerate. 
These results indicate that our method is effective for the growing use of personal-use LLM inference in resource-constrained single-GPU environments.

%
%
%
\bibliographystyle{splncs04}
\bibliography{sample}

\noindent{\footnotesize
This preprint has not undergone peer review (when applicable)
or any post-submission improvements or corrections.
The Version of Record of this contribution is published in
Euro-Par 2026: Parallel Processing,
and is available online at
\url{https://doi.org/10.1007/978-3-032-35251-4\_21}.
\par}

\end{document}